# Evaluation of *indirect* damage and damage saturation effects in dose-response curves of hypofractionated radiotherapy of early-stage NSCLC and brain metastases

Araceli Gago-Arias[1,2,3,*], Sara Neira[1], Miguel Pombar[2,4], Antonio Gómez-Caamaño[4,5], Juan Pardo-Montero[1,2,*]

1. Group of Medical Physics and Biomathematics, Instituto de Investigación Sanitaria de Santiago (IDIS), Santiago de Compostela, Spain.

2. Department of Medical Physics, Complexo Hospitalario Universitario de Santiago de Compostela, Spain.

3. Institute of Physics, Pontificia Universidad Católica de Chile, Santiago de Chile, Chile.

4. Group of Molecular Imaging and Oncology, Instituto de Investigación Sanitaria de Santiago (IDIS), Santiago de Compostela, Spain.

5. Department of Radiotherapy, Complexo Hospitalario Universitario de Santiago de Compostela, Spain.

**Corresponding authors:** Araceli Gago-Arias and Juan Pardo-Montero, Grupo de Física Médica e Biomatemáticas, Instituto de Investigación Sanitaria de Santiago (IDIS), Servizo de Radiofísica e Protección Radiolóxica, Hospital Clínico Universitario de Santiago, Trav. Choupana s/n, 15706, Santiago de Compostela (Spain); E-mails: maria.araceli.gago.arias@sergas.es, juan.pardo.montero@sergas.es; Phone: +34981955604

**Acknowledgements:** This project has received funding from the Instituto de Salud Carlos III (CPII17/00028 and PI17/01428 grants, FEDER co-funding). This project has received funding from the European Union's Horizon 2020 research and innovation programme under the Marie Skłodowska-Curie grant agreement No 839135.

**Abstract**

Background and purpose: To investigate the possible contribution of *indirect* damage and damage saturation to tumour control obtained with SBRT/SRS treatments for early-stage NSCLC and brain metastases.

Methods and Materials: We have constructed a dataset of early-stage NSCLC and brain metastases dose-response. These data were fitted to models based on the linear-quadratic (LQ), the linear-quadratic-linear (LQL), and phenomenological modifications of the LQ-model to account for indirect cell damage. We use the Akaike-Information-Criterion formalism to compare performance, and studied the stability of the results with changes in fitting parameters and perturbations on dose/TCP values.

Results: In NSCLC, a modified LQ-model with a beta-term increasing with dose yields the best-fits for $\alpha/\beta$=10 Gy. Only the inclusion of very fast accelerated proliferation or low $\alpha/\beta$ values can eliminate such superiority. In brain, the LQL model yields the best-fits, and the ranking is not affected by variations of fitting parameters or dose/TCP perturbations.

Conclusions: For $\alpha/\beta$=10 Gy, a modified LQ-model with a beta-term increasing with dose provides better fits to NSCLC dose-response curves. For brain metastases, the LQL provides the best fit. This might be interpreted as a hint of *indirect* damage in NSCLC, and damage saturation in brain metastases. The results for NSCLC are strongly dependent on the value of $\alpha/\beta$ and may require further investigation, while those for brain seem to be clearly significant. Our results can assist in the design of improved radiotherapy for NSCLC and brain metastases, aiming at avoiding over/under-treatment.

# Introduction

Stereotactic Body RadioTherapy (SBRT) and Stereotactic Radiation Surgery (SRS) constitute improvements in radiotherapy delivery **[1-3]**. Motivated by the increasing use of hypofractionation, there is an intense debate on the validity of the LQ-model **[4]** for large dose fractions **[5-9]**. Several investigators have shown that high-doses are less effective than would be expected from low-dose extrapolations **[10]**. Variations of the LQ-model were developed to describe this loss of relative effectiveness **[10-15]**, which we will refer to as *damage-saturation*. Differences in the effect of proliferation and hypoxia/re-oxygenation between conventional and hypofractionated schedules might also lead to different responses **[16,17]**. Recent results have also given rise to the hypothesis that high-doses trigger alternative mechanisms of cell death, sometimes called *indirect* cell death, related to vascular damage and immune-response activation **[18-26]**.

Indirect cell death effects seem to be well established experimentally, but there is controversy about their contribution to tumour control rates **[27,28]**. There is little evidence that indirect damage contributes to clinical tumour control rates obtained with SBRT/SRS: some studies have concluded that the LQ-model fits SBRT data adequately, and that the high levels of control achieved with SBRT are simply consistent with the level of dose escalation **[29,30]**; a recent analysis of outcomes of SBRT (lung cancer) and SRS (brain metastases) concluded that the LQ-model provides better fits to data than *damage saturation* models like the LQL and USC models **[31]**.

In this work we further explore the possible contribution of indirect cell death and/or damage saturation to tumour control obtained with SBRT/SRS for Non-Small-Cell-Lung-Cancer (NSCLC) and brain metastases (BM), by fitting different control models to dose-response curves. For the evaluation of models with various numbers of parameters we rely on the Akaike-Information-Criterion, a methodology widely used in radiotherapy.

## Materials and methods

**Clinical dataset**

We have analysed data for early-stage NSCLC and BM. We reviewed data included in references **[31,32]**, disposing of some schedules to increase consistency (see below), and included new data from articles published up to November 2018. Data was discarded when Kaplan-Meier local tumour control probabilities (TCP) were not reported, or could not be estimated from the information provided, when TCPs from very different fractionation regimens were not specified separately, and when reported TCP values did not correspond to the specified time points.

In total, 116 schedules were included in the database (61 for lung, 55 for brain). From the information provided in the publications, we collected: number of patients/metastases; schedule; dose prescription (defined as the average of isocentre and margin doses **[31,32]**); Kaplan-Meier local control (or local progression free survival/freedom from local relapse) reported at 1 and 3 years for brain and lung, respectively; gross tumour volumes (GTV). An overview of the schedules included in the database is presented in Table 1, showing the number of patients involved in the cohorts at different ranges of dose per fraction. Information about each treatment schedule is summarised in Tables SM1 and SM2. Further information is provided in a datasheet available from **[33]**.

**Radiobiological models**

The LQ-model **[4,34]** has been used as the reference model in this study. The surviving fraction of a population of cells after a dose *d* can be expressed as,

$$\log SF_{LQ} = -\alpha d - \beta d^2 \qquad (1)$$

Among models that consider damage saturation, we have restricted our analysis to the LQL model:

$$\log SF_{LQL} = -\alpha d - \frac{2\beta}{\delta^2}\left(\delta d + e^{-\delta d} - 1\right) \quad (2)$$

There have been attempts at modelling the contribution of *indirect* damage **[35-37]**, but there is no simple closed-form expression. As suggested in **[37]**, we have investigated *ad hoc* modifications of the LQ-model with dose-dependent alpha-beta terms of the following form:

$$\log SF_{mod} = -\alpha d\left(1 + ad^{a'}\right) - \beta d^2\left(1 + bd^{b'}\right) \quad (3)$$

where *a'* and *b'* take values of 1/2, 1/3, 1 and 2, in order to explore a variety of soft to strong dose dependencies. Note that single-term extensions of either the linear or quadratic terms can be investigated by setting *b* or *a* to zero. Several biological processes may be associated with the proposed modification: i) negative *a* and *b* values might be caused by lack of reoxygenation with hypofractionation; ii) negative *b* values might be associated with increasing damage repair at high-doses, as in the LQL; and iii) positive values of *a* and *b* might be related to indirect cell death (especially for *b*, if such mechanisms are predominant at large doses). Accordingly, we have set *a*≤0 and no specific sign for *b* in our study.

For a treatment of *n* fractions the overall surviving fraction at the end of the treatment can be written as **[38]**,

$$SF_{overall} = \left(\prod_{i=1}^{n} SF_i\right) \exp\left(\lambda \, max\left(0, T - T_k\right)\right) \quad (4)$$

where $SF_i$ is the surviving fraction of each radiation fraction (modelled with any of the models presented above), and the exponential accounts for accelerated proliferation when treatment time *T* exceeds the proliferation kick-off time $T_k$.

**TCP models**

To quantify treatment outcome, tumour control probability was modelled using a logistic function

[32,39],

$$TCP = \frac{TCP_{max}}{1 + \left(\frac{D_{50}}{EQD2_{model}}\right)^{4\gamma_{50}}}$$ (5)

where $TCP_{max}$ sets an upper TCP limit, $D_{50}$ is the dose corresponding to 50% control and $\gamma_{50}$ is the normalised dose-response gradient. Model-derived $EQD2_{model}$ values were calculated using the different models under investigation. The normalised dose-response gradients were fixed to $\gamma_{50}$=0.83 and 0.7 for NSCLC and BM, respectively, according to [40].

We also explored the Poissonian formulation of TCP with population-averaging of radiosensitivities [41,42], by using a probability density function for the distribution of the parameter $\alpha$:

$$TCP = TCP_{max} \cdot \int \exp\left(-N_0 SF_{overall}(\alpha)\right) f(\alpha) d\alpha$$ (6)

where $N_0$ is the number of tumour clonogens, and $SF_{overall}$ is the surviving fraction given by Eq. (4). We used the gamma distribution for the averaging of radiosensitivities as in [31]. $N_0$ was set to $10^5$, and the shape parameter of the gamma distribution, $g$, was set to match the normalised dose-response gradients presented above (0.83 for NSCLC, 0.7 for brain).

**Evaluation of goodness-of-fit**

The Akaike-Information-Criterion value with sample size correction for a given model ($AIC_c$) is given by [43,44]:

$$AIC_c = -2\log(L) + 2k + 2k(k+1)/(N_s - k - 1)$$ (7)

where $k$ is the number of model parameters, $N_s$ the sample size, and $L$ the likelihood function for the model:

$$L = \prod_i f_b\left(N_{s,i} \times TCP_{model}, N_{s,i}, c_i\right) \tag{8}$$

where $i$ spans the number of points included in the dataset (treatment cohorts), and $N_{s,i}$ is the number of patients in each cohort. $f_b$ denotes the binomial probability function (the probability of achieving $N_{s,i} \times TCP_{model}$ controls in $N_{s,i}$ patients if the experimental control probability is $c_i$). In order to derive realistic estimates for the experimental tumour control probabilities, $c_i$, including the sample size of every cohort, the Agresti–Coull approximation was used, and confidence intervals were calculated with the Clopper-Pearson method [45,46]. This provides an estimation of the most likely TCP value (probability of success of a Bernoulli trial) from the Kaplan-Meier TCP value and the number of patients involved in the study.

In total, 26 models were analysed (LQ, LQL and 24 combinations of different functional dependencies in the modified LQ model). Notice that the model with $a'=1$ and $b'=0$ would be equivalent to a LQ model with an effective $\beta$ parameter value $\beta_{eff}=a\alpha+\beta$. The model exhibiting the lowest $AIC_c$ is considered the best-fitting model. The difference between the $AIC_c$ value of the LQ-model, taken as reference, and the $AIC_c$ of the another model $j$ is defined as $\Delta AIC_{c,j}=AIC_{c,LQ}-AIC_{c,j}$. We also compute evidence ratios, $EVR$, for each model (see Supplementary Materials). Moreover, log-likelihood ratio tests were performed, obtaining $p$-values to check the significance of the results.

A simulated annealing method [47] was implemented to maximize the likelihood. For the fit to the logistic model we allowed some parameters to be free ($D_{50}$, $\lambda$, and $\delta$, $a$ and $b$), while fixing values for some others ($\alpha$, $\alpha/\beta$, $T_k$, $\gamma_{50}$, and $TCP_{max}$). The LQ-model radiosensitivity parameter $\alpha$ was given a fixed value of 0.3 Gy$^{-1}$, while $TCP_{max}$ was set to 0.95 [32]. We also set $\alpha/\beta=10$ Gy, and $T_k=28$d, consistent with the range of reported values [32,48,49], but other values were also explored. For Poisson-TCPs the radiosensitivity parameter $\alpha$ was optimised instead of $D_{50}$.

**Variation of parameters and dose/response perturbations**

We have investigated whether using different values of the $\alpha/\beta$ ratio, the proliferation kick-off time,

and the slope of the TCP curve ($\gamma_{50}$/$g$ for logit/Poisson, respectively) would affect our results and conclusions.

We also studied the sensitivity to uncertainties in dose and TCP values, the aim being to explore whether the rank/performance of models are affected by inherent experimental uncertainties. Dose and/or TCP values were perturbed, and the optimization process and $AIC_c$ analysis performed with the perturbed dataset. This procedure was repeated 20 times (50 times when dose and TCP values were simultaneously perturbed). For dose perturbations, we chose a normal distribution with a mean equal to the dose per fraction and a 5% relative standard deviation. Tumour control probability values were sampled from the binomial distribution, using the Agresti-Coull probability associated with each cohort and the number of patients involved.

**Implementation**

The code and data in Matlab format (Mathworks, Natick, MA) are available from **[33]**.

# Results

The study of NSCLC dose response data shows improved fit quality for the modified LQ-model, compared with the LQ. This trend was observed both with the logistic and Poisson formulations. The modified LQ-model with a square root dependency of the dose in the beta term ($b'$=1/2), provides the best results with $b>0$ ($b$=0.10 and $b$=0.06 for logit and Poisson fits, respectively), which implies an increasing $\beta$-term with increasing dose. Consistently with this, the LQL model, which describes a *decreasing* effectivity with increasing dose, was not able to improve the quality of the fit achieved with the LQ model. In the top panels of Figure 1 we present best-fits to TCP data with the LQ and modified LQ models for the logistic TCP methodology, and in Table 2 we report $AIC_c$ and EVR values. Although the fitting improvement cannot easily be perceived on visual inspection, the $AIC_c$ supports the difference in fit quality, with $\Delta AIC_c$~6 and EVR~20 ($p$=0.004) for the modified LQ model with b'=1/2.

Further information is provided in the Supplementary Materials: in Figures SM1 and SM2 we present best fits (TCP vs $EQD_2$) for each model; best-fitting parameters in Table SM3, and; AICs and EVRs in Table SM4. This information is included for the sake of completeness and reproducibility.

Analysis of BM data shows that the LQL provides the best-fit. The results obtained with the LQL model are very similar to those obtained with the modified LQ-model with $b'=1/2$ and $b<0$ ($b=-0.13$ for logit and Poisson). This is not unexpected as this model (any modified LQ with $b<0$) behaves similarly to the LQL model. In the bottom panels of Figure 1 we present best-fits for those models, with the logit methodology.

While the improvement in the fit is again hardly noticeable in the figure, $\Delta AIC_c$ of the best model (the LQL) is far larger than in the case of NSCLC, $\Delta AIC_c \sim 125$, which results in values of $EVR>10^{20}$ ($p<10^{-11}$) as shown in Table 2.

We report complementary information in the Supplementary Materials: in Figures SM3 and SM4 we present best fits (TCP vs $EQD_2$) for each model; AICs and EVR values for each model in Table SM5, and; best-fitting parameters for each model in Table SM6.

Regarding the study of parameter variation effects, in Table 3 we present $AIC_c$ and EVR values for the best-fitting models (modified LQ-model with $b'=1/2$ for NSCLC and brain, and LQL for brain) and the LQ-model, when different fixed values are set for parameters $T_k$, $\alpha/\beta$ and $\gamma_{50}$. Results reported in this table correspond to the logistic TCP methodology.

In NSCLC, the sooner the start of accelerated proliferation (the lower value of $T_k$), the lower the evidence ratio of the modified LQ-model. $\Delta AIC_c$ values decrease from 6 (EVR~21 and $p=0.004$) for $T_k=28d$ to 1.4 (EVR~2 and $p=0.06$) for $T_k=7d$. When changing the value of $\gamma_{50}$, which models the slope of the dose-response curve, the superiority of the modified LQ over the LQ-model diminishes with increasing $\gamma_{50}$ (steeper slopes), and totally disappears when $\gamma_{50}=1.5$. For NSCLC, the effect of the $\alpha/\beta$ value has been explored in detail. For $\alpha/\beta \geq 10$ Gy the modified LQ with $b>0$ is significantly superior to the LQ (EVR~$2\times10^8$ and $p<10^{-11}$ for $\alpha/\beta=20$ Gy; EVR~21 and $p<0.005$ for $\alpha/\beta=10$ Gy). This significance disappears at $\alpha/\beta$~8 Gy, where the LQ model can be considered the best. If the $\alpha/\beta$ is further decreased, for $\alpha/\beta$~5Gy the behaviour observed at large $\alpha/\beta$ values is inverted, and models with decreasing radiosensitivity with increasing dose (modified LQ with $b<0$ and LQL, the latter not shown in Table 3) become superior to the LQ ($p<0.01$).

On the other hand, for BM the superiority of the LQL over the LQ model is not importantly affected by variations of $T_k$, $\alpha/\beta$ and $\gamma_{50}$, with values of $\Delta AIC_c>11$, EVR>289 and $p<3\times10^{-4}$ for every investigated set of parameters.

Identical trends are observed when using the Poisson methodology, both for NSCLC and BM, as shown in Tables SM7 and SM8.

The quality of the fits changes drastically when dose/TCP values are perturbed, as can be observed in the mean and standard deviation of $AIC_c$ values for the LQ-model reported in Table 4. Accordingly, EVR and $\Delta AIC_c$ values suffer large variations, but the modified LQ model (for NSCLC) and LQL (for BM) remain systematically better than the LQ model in spite of dose/TCP perturbations.

In the case of NSCLC, the modified LQ model is superior to the LQ model in 20 out of 20 experiments with dose perturbations, 17 out of 20 with TCP perturbations, and 33 out of 50 experiments with both TCP and dose perturbations. A z-test of proportions shows that the superiority of the former model is significant, with $p<10^{-5}$, $p<10^{-3}$, and $p=0.01$, respectively. In the case of BM, dose/TCP perturbations do not affect at all the rank of the LQL model, which is superior in every single experiment with dose, TCP, and dose and TCP perturbations ($p<10^{-5}$).

In Figure 2 we show total dose versus dose per fraction isoeffective curves calculated with the different models for two $\alpha/\beta$ values. This intends to illustrate the clinical implications of the LQ model modifications, which improves the description of experimental dose response data, with respect to the LQ. We chose $\alpha/\beta$=10 Gy, the value conventionally accepted, and $\alpha/\beta$=5 Gy, which provided better fit quality indicators for both NCLCL and BM. Differences between isoeffect curves calculated with the different models are subtle for NSCLC, while for brain the loss of effectiveness with increasing dose per fraction described by the LQL leds to important discrepancies with the LQ model.

## Discussion

We have analysed the dose-response for NSCLC and brain metastases, looking for evidence of *indirect* damage and/or damage saturation. The methodology that we followed consisted of fitting dose-response curves to response models based on the conventional LQ-model, the LQL model, and simple modifications of the LQ-model to allow for increasing/decreasing relative radio-sensitivities with increasing dose. If mechanisms accounting for decreasing radio-sensitivity with increasing dose are dominant (like damage saturation or lack of re-oxygenation) the LQL should prove superior; if contribution of indirect damage with increasing dose is dominant, the modified LQ should prove superior.

For NSCLC, a modified LQ model with a slow increase of beta with dose ($\sim\sqrt{d}$) provides the best fit ($\Delta AIC_c\sim6$, EVR~20, $p$=0.004). This might be interpreted as a contribution of indirect cell damage at large doses, as observed in experimental studies **[8,6,19]**. On the other hand, the LQL model provided the best fit for BM.

There are several confounding factors that can affect these conclusions. For example, if accelerated proliferation kicks off in a relatively short time, it would create an effect similar to that observed in NSCLC: treatments with few fractions will not be affected by proliferation, but conventional treatments will, creating an effect similar to an increasing relative radiosensitivity with increasing fraction size. In fact, we have observed that by decreasing $T_k$ from the typical 28 day value, the statistical superiority of the modified LQ-model also decreases. However, we need to reach values of $T_k\sim7$ day in order to eliminate the superiority of the modified LQ-model. Such fast kick-off times are not consistent with reported values **[32,48]**.

Different $\alpha/\beta$ ratios also affect the fits. While for BM, values of $\alpha/\beta$ in the range of 5-20 Gy do not affect the conclusions (the LQL being superior to the LQ model), the behaviour for NSCLC is different, and particularly interesting. For $\alpha/\beta \geq 10$ Gy the modified LQ with $b>0$ is significantly superior to the LQ, the level of significance increasing with the value of $\alpha/\beta$. This trend disappears at $\alpha/\beta \sim 8$ Gy ($p=0.13$), where the LQ model can be considered the best. If the $\alpha/\beta$ is further decreased, for $\alpha/\beta \sim 5$ Gy the effect found at large $\alpha/\beta$ values is inverted, and models with decreasing radiosensitivity with increasing fraction size (modified LQ with $b<0$ and LQL) become superior to the LQ. Recent estimations of the $\alpha/\beta$ parameter for early-stage NSCLC from clinical dose response data place its value at around 13 Gy **[49]**, and in vitro studies have also reported $\alpha/\beta$ ratios above 10 Gy for different lung tumour cell lines **[50]**. However, a recent work reported lower values ($\alpha/\beta \sim 3$ Gy) using a complex mechanistic model **[32]**, while works analysing the clinical reponse of advanced NSCLC have also provided low $\alpha/\beta$ values (~4 Gy) [51]. The hypothesis of $\alpha/\beta<10$ Gy for NSCLC and the LQ or LQL being the best-fitting models should not be completely discarded.

The heterogeneity of reported endpoints can be another source of uncertainty. For example, local control may be assesed differently among studies, and that information may not always be provided. More importantly, if the analysis includes endpoints that consider local control and death or only local control as events, it may biased. A detailed analysis of the papers in our dataset showed that the reported endpoints do not include death as an event. We have also shown that the conclusions of this work stand under dose/response perturbations. Nonetheless, special care should be put on considering homogeneous endpoints to avoid potential issues.

We have also evaluated whether differences in tumour volumes may introduce a bias: if small NSCLC tumours are prescribed larger doses per fraction than large tumours, it could be interpreted as an increasing relative radiosensitivity with increasing doses, because small tumours are easier to control. The opposite should happen in BM in order to explain the superiority of the LQL model. We have analysed reported GTVs, finding no association between volume and prescribed dose-per-fraction: neither a Spearman test of GTV-dose nor a hypothesis-test of tumours below $V_t$ receiving higher doses showed significance (with volumes $V_t$ set to the median of the tumour volumes amongst all the cohorts, 25 cm$^3$ and 2 cm$^3$ for lung and brain, respectively), see Figure SM5.

A recent study [31] found that the LQ-model provided a better fit to dose-response curves for NSCLC and BM than the LQL model. Our results are in disagreement with this study for BM. We attribute part of the disagreement to the management of the averaging of radiosensitivities, as Shuryak *et al.* included what is called intra-tumour averaging in the computation of Poisson-TCPs, which cannot decrease the slope of the response curve as the inter-patient averaging does. Moreover, Shuriak *et al.* did not account for the effect of tumour proliferation and assumed a maximum TCP value equal to 1 instead of the 0.95 used here. We have compared our fitting methodology to Shuryak's methodology on our BM dataset, and found that better fits are obtained with the former (e.g. AIC=885 versus 1450, for the LQ model). On the other hand, all models tested in [31] are models with decreasing radiosensitivity with increasing dose. Therefore, Shuryak *et al.* could not detect the effect of increasing radiosensitivity with increasing dose in NSCLC that we have found in this work.

The significance of the differences among models can be questioned. The likelihood ratio test provides $p$ values below 0.05 for $\Delta AIC_c > 2.3$. The literature on AIC usually sets stronger significance thresholds: models with $\Delta AIC_c < 2$ are generally considered to be equally good, while for values between 2–6 models are rarely dismissed. Model rejection is usually considered when $\Delta AIC_c > 6$, or 9-11 in more conservative approaches [52,53,31]. For an $\alpha/\beta$ ratio equal to 10 Gy, the modified LQ-model for NSCLC has $\Delta AIC_c \sim 6$ with $p=0.004$. On the other hand, the LQL model for BM has $\Delta AIC_c \sim 125$ with $p<10^{-11}$, a result that should be considered significant. However, it is important to point out that fits to brain data are significantly worse than fits to NSCLC data: $AIC_c \sim 700$ vs. 350. This is probably caused by the larger heterogeneity of BM. Such heterogeneities might affect the significance of the superiority of the LQL model.

Our results should stimulate further research in this area (especially for NSCLC due to the observed behaviour with $\alpha/\beta$ and conflicting reports on the value of this parameter), and may assist with the design of optimal radiotherapy treatments for NSCLC and BM, aiming at avoiding over- or under-treatment. Dose prescription to such tumours might be reconsidered in the light of the reported evidence. Assuming that the $\alpha/\beta$ is ≥10 Gy, our results suggest that, a) prescription doses for extremely hypofractionated early-stage NSCLC treatments might be decreased to reduce toxicity without strongly compromising control and, b) extremely hypofractionated treatments should be avoided for BM. The response of normal tissues to such doses must always be considered in order to properly define the therapeutic window.

**FIGURES AND TABLES**

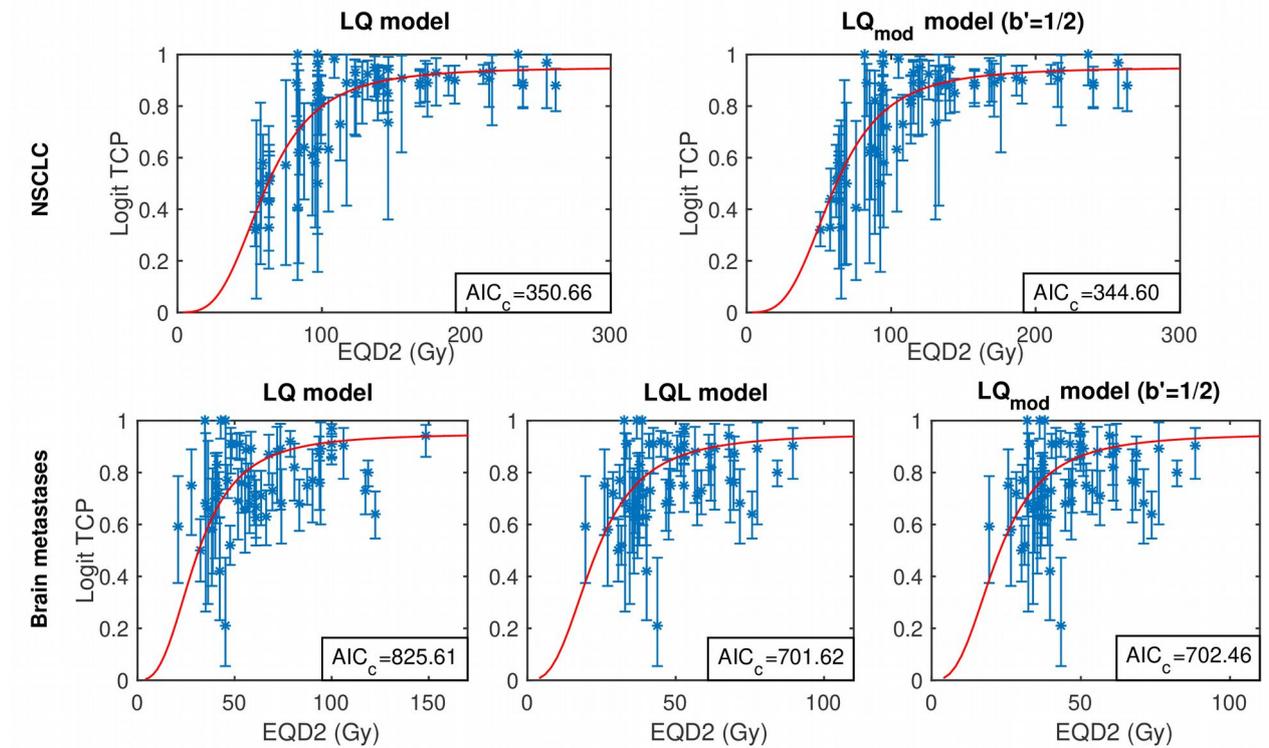

Figure 1: Best fits to early-stage NSCLC (top panels) and brain metastases (bottom panels) dose response data. Results are shown for the logistic TPC formulation using the LQ model (top and bottom left), the LQ modification with functional dependency $b'=1/2$ (right), and the LQL model (center bottom).

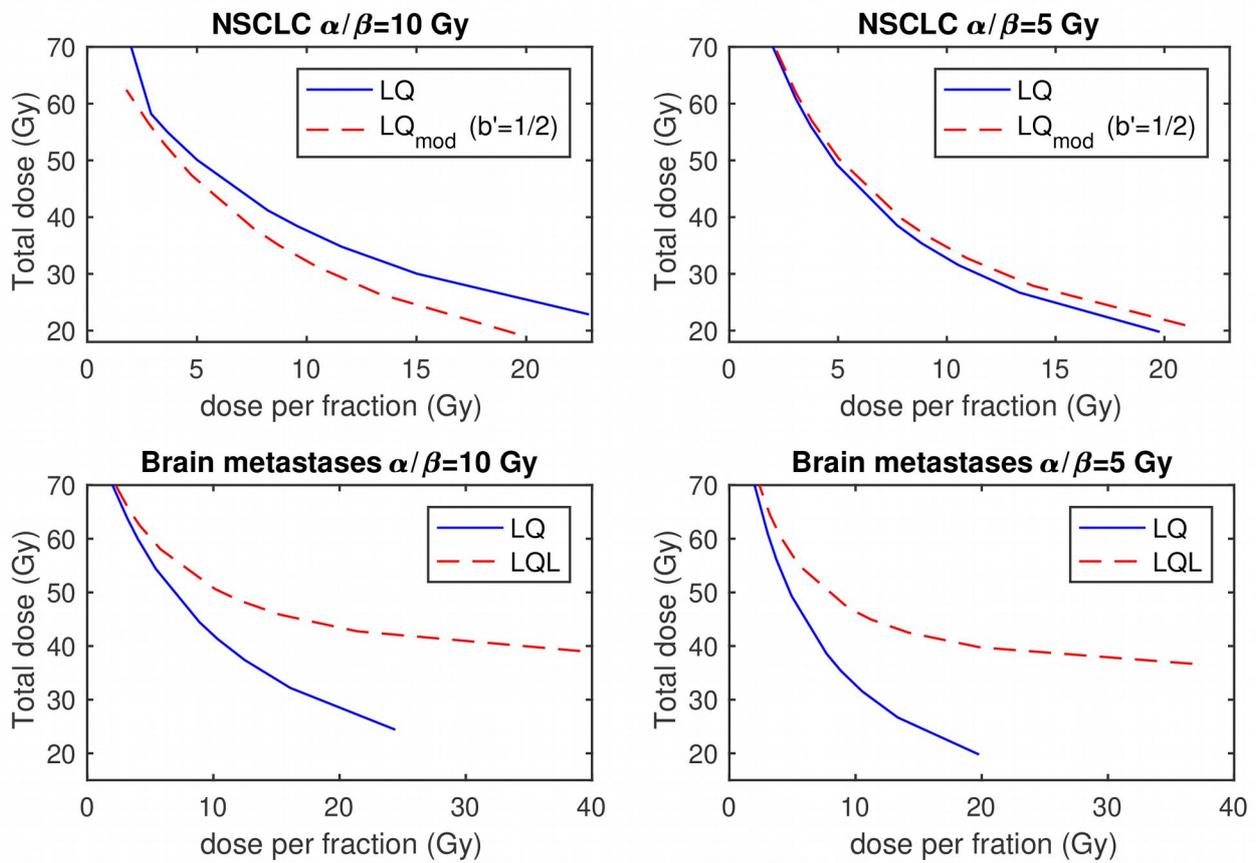

Figure 2: Total isoeffective dose vs dose per fraction for the LQ model, the modified LQ with b'=1/2 for NSCLC (top panels), and the LQL model for brain metastases (bottom panels), with $\alpha/\beta$=10 Gy (left) and 5 Gy (right), respectively. The isoeffect corresponds to that of a conventional treatment with 70 Gy in 35 fractions with weekend breaks calculated with the LQ model.

| Treatment site | Dose per fraction range (Gy) | Median dose per fraction* (Gy) | # pts | #cohorts** | Median total dose* (Gy) | Total dose range (Gy) |
|---|---|---|---|---|---|---|
| NSCLC | [1.8-2.7 ] | 2.0 | 940 | 8 | 62.6 | [48.8-94.2] |
|  | [3.8-9.9 ] | 8.0 | 188 | 10 | 57.0 | [36.0-75.0] |
|  | [10.4-14.4 ] | 12.6 | 1144 | 20 | 49.0 | [37.6-70.0] |
|  | [15.6-19.9 ] | 17.9 | 547 | 8 | 53.7 | [46.6-59.7] |
|  | [20.2-24.9 ] | 22.3 | 615 | 11 | 64.1 | [20.2-70.9] |
|  | [27.0-33.7] | 28.1 | 107 | 4 | 28.1 | [27.0-33.7] |
| Brain Metastases | [3.9-4.9 ] | 4.9 | 97 | 4 | 36.6 | [29.3-48.6] |
|  | [5.1-9.8 ] | 6.2 | 952 | 20 (1) | 30.3 | [21.6-64.7] |
|  | [10.8-15.0 ] | 13.3 | 159 | 6 (2) | 49.4 | [38.0-68.6] |
|  | [16.6-19.4 ] | 18.0 | 412 | 5 (1) | 19.1 | [17.0-46.6] |
|  | [20.2-24.3 ] | 22.5 | 1231 | 12 (2) | 22.8 | [20.3-52.5] |
|  | [26.1-37.5] | 28.9 | 885 | 8 (1) | 30.0 | [26.1-63.8] |

\* Median of the cohorts, with no consideration of the number of patients involved
\*\* Data in parentheses specify the number of cohorts with adjuvant whole brain radiotherapy

Table 1: Overview of the databases of NSCLC and Brain Metastases considered for the study.

|  | NSCLC | | | | Brain Metastases | | | |
|---|---|---|---|---|---|---|---|---|
|  | Logit model | | Poisson model | | Logit model | | Poisson model | |
|  | $AIC_c$ | EVR | $AIC_c$ | EVR | $AIC_c$ | EVR | $AIC_c$ | EVR |
| **LQ** | 350.66 | 1.00 | 358.53 | 1.00 | 825.61 | 1.00 | 844.75 | 1.00 |
| **LQL** | 352.76 | 0.35 | 359.72 | 0.55 | 701.62 | 8.4E+26 | 709.63 | 2.2E+29 |
| **LQmod (a'=1/2)** | 352.87 | 0.33 | 360.41 | 0.39 | 827.85 | 0.33 | 846.45 | 0.43 |
| **LQmod (a'=1)** | 352.88 | 0.33 | 360.24 | 0.43 | 788.43 | 1.2E+08 | 804.11 | 6.7E+08 |
| **LQmod (b'=1/2)** | 344.60 | 20.68 | 352.67 | 18.70 | 702.46 | 5.5E+26 | 711.91 | 7.0E+28 |
| **LQmod (b'=1)** | 348.65 | 2.72 | 356.69 | 2.51 | 703.68 | 3.4E+10 | 711.38 | 9.1E+28 |
| **LQmod (a'=1/2, b'=1/2)** | 346.91 | 6.52 | 354.98 | 5.88 | 704.75 | 1.7E+26 | 713.17 | 3.7E+28 |
| **LQmod (a'=1/2, b'=1)** | 350.92 | 0.88 | 358.77 | 0.89 | 705.98 | 9.5E+25 | 713.41 | 3.3E+28 |
| **LQmod (a'=1, b'=1/2)** | 346.86 | 6.68 | 354.95 | 5.98 | 704.75 | 1.7E+26 | 712.78 | 4.5E+28 |
| **LQmod (a'=1, b'=1)** | 350.95 | 0.87 | 358.70 | 0.92 | 706.04 | 9.2E+25 | 713.33 | 3.4E+28 |

Table 2: Akaike-information-criterium ($AIC_c$) and evidence ratio (EVR) values for the fits of the logit/Poisson TCP models (based on the LQ model, taken as reference, the LQL model, and modifications of the LQ models) to the early stage NSCLC and brain metastases datasets.

|  | NSCLC | | | | Brain Metastases | | | | | | |
|---|---|---|---|---|---|---|---|---|---|---|---|
|  | AIC$_c$ LQ | AIC$_c$ LQmod b'=1/2 | EVR (LQmod b'=1/2) | b (LQmod b'=1/2) | AICc LQ | AICc LQL | AICc LQmod b'=1/2 | EVR (LQL) | δ (LQL) | EVR (LQmod b'=1/2) | b (LQmod b'=1/2) |
| $T_k$ **variation** | | | | | | | | | | | |
| $T_k$ = 28 d | 350.7 | 344.6 | 20.7 | 0.10 | 825.6 | 701.6 | 702.5 | 8E+26 | 0.14 | 6E+26 | -0.13 |
| $T_k$ = 21 d | 348.4 | 344.4 | 7.3 | 0.10 | 825.6 | 701.7 | 702.5 | 8E+26 | 0.14 | 6E+26 | -0.13 |
| $T_k$ = 14 d | 345.4 | 343.6 | 2.6 | 0.06 | 825.6 | 701.7 | 702.4 | 8E+26 | 0.14 | 6E+26 | -0.13 |
| $T_k$ = 7 d | 345.3 | 343.9 | 2.0 | 0.07 | 825 | 679.7 | 676.2 | 4E+31 | 0.18 | 2E+32 | -0.14 |
| $\alpha/\beta$ **variation** | | | | | | | | | | | |
| $\alpha/\beta$ = 20 Gy | 420.3 | 350.1 | 2E+15 | 0.50 | 752.2 | 709.4 | 711.6 | 2E+09 | 0.09 | 7E+08 | -0.11 |
| $\alpha/\beta$ = 10 Gy | 350.7 | 344.6 | 20.68 | 0.10 | 825.6 | 701.6 | 702.5 | 8E+26 | 0.14 | 6E+26 | -0.13 |
| $\alpha/\beta$ = 8 Gy | 341,2 | 341.2 | 1.0 | 0.05 | - | - | - | - | - | - | - |
| $\alpha/\beta$ = 6 Gy | 335.1 | 337.3 | 0.3 | 9.7E-05 | - | - | - | - | - | - | - |
| $\alpha/\beta$ = 5 Gy | 339.7 | 335.3 | 8.9 | -0.03 | 917.1 | 694.5 | 696.6 | 2E+48 | 0.21 | 8E+47 | -0.13 |
| $\gamma_{50}$ **variation** | | | | | | | | | | | |
| $\gamma_{50}$ = 1.5 | 369.7 | 371.9 | 0.3 | 8.7E-03 | - | - | - | - | - | - | - |
| $\gamma_{50}$ = 1 | 347.5 | 346.5 | 1.6 | 0.05 | 963.6 | 776.2 | 766.3 | 5E+40 | 0.19 | 7E+42 | -0.14 |
| $\gamma_{50}$ = 0.83 | 350.7 | 344.6 | 20.7 | 0.10 | 888.4 | 732.9 | 731.8 | 6E+33 | 0.17 | 1E+34 | -0.13 |
| $\gamma_{50}$ = 0.7 | 355.2 | 344.9 | 170.8 | 0.24 | 825.6 | 701.6 | 702.5 | 8E+26 | 0.14 | 6E+26 | -0.13 |
| $\gamma_{50}$ = 0.3 | - | - | - | - | 658.3 | 647.0 | 647.3 | 289.0 | 0.08 | 2E+02 | -0.07 |

Table 3: Effect of variations of fitting parameters $T_k$, $\alpha/\beta$ and $\gamma$ on the evidence ratio (EVR) of the modified LQ model with b'=1/2 (NSCLC and brain metastases) and the LQL (brain metastases), relative to the LQ model. AIC$_c$ values for the LQ model fit are reported to show fit improvements with parameter variations. $b$ and $\delta$ parameter values of the modified LQ and LQL models respectively are also included. Parameters that were not varied were set to $T_k$ = 28 d, $\alpha/\beta$ = 10 Gy and $\gamma_{50}$ = 0.83 (NCSLC) or 0.7 (brain).

|  | NSCLC | | | Brain Metastases | | |
|---|---|---|---|---|---|---|
|  | AIC$_c$ LQ | ΔAIC$_c$ LQmod b'=1/2 | EVR (LQmod b'=1/2) | AICc LQ | ΔAIC$_c$ LQL | EVR (LQL) |
| **Dose Perturbations** | 375 ± 17 | 11.4 ± 6.5 | 8E+05 ± 3E+06 | 818 ± 31 | 119.5 ± 18.1 | 2E+31 ± 8E+31 |
| **TCP perturbations** | 384 ± 21 | 2.2 ± 3.1 | 36 ± 147 | 832 ± 50 | 114.2 ± 23.7 | 1E+35 ± 5E+35 |
| **Dose & TCP perturbations** | 409 ± 29 | 4.2 ± 6.6 | 2479 ± 9163 | 849 ± 51 | 119.6 ± 30.4 | 2E+39 ± 1E+40 |

Table 4: Mean ± one standard deviation of AIC$_c$, ΔAIC$_c$ and EVR values for fits with the LQ model, LQL model (for brain metastases) and modified LQ model with b'=1/2 (for NSCLC) when doses, tumor control probabilities, or both values in the dataset were perturbed. Fits were performed with the logistic TCP formulation using $T_k$ = 28 d, $\alpha/\beta$=10 Gy and $\gamma_{50}$ =0.83 (NSCLC) or 0.7 (brain).